\documentstyle[12pt]{article}
\newlength{\extraspace}
\newlength{\extraspaces}
\begin{document}
\addtolength{\baselineskip}{.7mm}

\thispagestyle{empty}

\begin{flushright}
{\sc KAIST-CHEP}-95/18\\
May 1996
\end{flushright}
\vspace{.3cm}

\begin{center}
{\Large\bf{Particle Dynamics in a Class of 
 \\[2mm]   2-dimensional Gravity Theories 
}}\\[10mm]
{\sc Daesung Hwang and Youngjai Kiem}\\[3mm]
{\it Department of Physics\\[2mm]
 Sejong University\\[2mm]
Seoul 143-747, Korea\\[2mm] }
{\sc And}\\[3mm]
{\sc Dahl Park}\\[3mm]
{\it Department of Physics\\[2mm]
 KAIST\\[2mm]
Taejon 305-701, KOREA\\[5mm] }
{e-mail: dshwang@phy.sejong.ac.kr \\
        ykiem@phy.sejong.ac.kr \\
        dpark@chep6.kaist.ac.kr }
\\[15mm]

{\sc Abstract}
\end{center}

We provide a method to determine the motion of a classical massive 
particle in a background geometry of 2-dimensional gravity theories,
for which the Birkhoff theorem holds.  In particular, we get the 
particle trajectory in a continuous class of 2-dimensional dilaton 
gravity theories that includes the Callan-Giddings-Harvey-Strominger 
(CGHS) model, the Jackiw-Teitelboim (JT) model, and the $d$-dimensional 
$s$-wave Einstein gravity.  The explicit trajectory expressions for 
these theories are given along with the discussions on the results.
\noindent

\vfill

\newpage

\section{Introduction}

The complexity of 4-dimensional general relativity encountered in its 
analytical treatment is considerable.  After long and varied attempts, 
many interesting quantum and classical questions on the gravitation
still resist their analytical solutions.  One motivation for the 
development of 2-dimensional gravity theories in recent years 
\cite{banks} \cite{giddings} is to search for the
model theory that captures the essential features of 4-dimensional
general relativity
and at the same time provides us with a manageable framework of 
analysis.  The big reduction of the degrees of freedom in the gravity
sector of a 2-dimensional theory is obvious, while it is plausible that the
$s$-wave sector of a 4-dimensional theory is, at least at a superficial
level, 2-dimensional and thereby shares some common physical properties
with 2-dimensional gravity theories, such as Jackiw-Teitelboim
(JT) model \cite{jt} and Callan-Giddings-Harvey-Strominger (CGHS) model
\cite{cghs}.  The developments in quantum gravitational issues,
including the discussions of the black hole information paradox, afforded 
by these model theories are by now reported in literature \cite{giddings}.

The modest aim of this paper is to better understand the classical 
dynamics of a class of 2-dimensional dilaton gravity theories, which
includes many theories of interest as its special cases.  
In particular, we analytically determine the motion of a point 
massive particle in the background space-time 
geometry of the 2-dimensional dilaton gravity theories.  The gravity 
sector action in our consideration is thus given by
\begin{equation}
I_g = \int d^2 x \sqrt{-g} e^{-2 \phi} [\, R+ \gamma g^{\alpha \beta}
\partial_{\alpha} \phi \partial_{\beta} \phi+ \mu e^{2\lambda\phi}\, ],
\label{a1}
\end{equation}
where $R$ and $\phi$ represent the 2-dimensional scalar curvature and
the dilaton field, respectively. 
The parameters $\gamma$, $\mu$ and $\lambda$
are arbitrary real numbers.  A specific choice of
them corresponds to a particular gravity theory, as shown in the table.
\begin{equation}
\begin{tabular}{|c|c|c|c|} \hline
 &$\gamma$ & $\lambda$ &
$a=2-\lambda-\gamma /4$ \\ \hline
JT & 0   & 0 & 2   \\
CGHS & 4 &  0 & 1 \\
$d$-D Einstein $(3<d<\infty )$ & $4{d-3\over d-2}$ &
${2 \over d-2}$ & $0< {d-3 \over d-2} < 1$ \\ \hline
\end{tabular}
\label{table}
\end{equation}
Among many theories of the form (\ref{a1}), three cases have
been under the most intensive scrutiny.  First, in case of the JT
model, one of the earliest models to realize non-trivial gravitational 
dynamics in 2-dimensional space-time, many properties and reformulations
are available in literature \cite{jt} \cite{otherjt}.  
Second, the string-theory-inspired CGHS model
provided us with an analytically tractable framework for the (quantum)
study of the gravitation \cite{giddings}.  Another case of importance is
the spherically symmetric reduction of the $d$-dimensional 
Einstein-Hilbert action \cite{reduc} \cite{birkhoff}. In this case,
the 2-d dilaton field $\phi$ is directly related to the geometric radius
of the $(d-2)$-dimensional sphere in the $d$-dimensional spherically 
symmetric geometry.  To be specific, we write the
spherically symmetric $d$-dimensional metric $g^{(d)}_{\mu \nu}$
as the sum of longitudinal and transversal parts 
\begin{equation}
 ds^2 = g_{\alpha \beta} dx^{\alpha}dx^{\beta} - 
\exp { ( - \frac{4}{d-2} \phi)} d \Omega^2 ,
\label{a3}
\end{equation}
where $d\Omega^2$ is the sphere $S^{d-2}$ of unit radius, 
and we use the metric signature $(+- \cdots -)$.
The $d$-dimensional Einstein-Hilbert action 
\begin{equation}
 I = \int d^d x\, \sqrt{-g^{(d)}}\, R^{(d)} ,
\label{a4}
\end{equation}
where $R^{(d)}$ is the $d$-dimensional scalar curvature,
reduces to (\ref{a1}) with the specified
values of parameters in the table (\ref{table}).

The study of the classical particle trajectory in each model, which
is the main focus of this paper, is a good way to (partially) understand
the relationship among the theories described by (\ref{a1}) and,
eventua1ly, to gain more understandings of the Einstein gravity,
its relation to string-motivated gravity, and peculiarities
of each theory.  To be specific, we will obtain the particle 
trajectory $x(\tau )$, i.e., the solution of the geodesic equation 
resulting from the point particle action
\begin{equation}
I_m = -m\int ds
=-m\int d^2x \int d\tau \delta^{(2)}(x-x(\tau))[\, g_{\alpha \beta}
\dot{x}^{\alpha}(\tau) \dot{x}^{\beta}(\tau)\, ]^{1/2}
\label{a2}
\end{equation}
in each theory.  In case of the JT model, the same problem was solved by 
the authors of Ref. \cite{hksy}.  The main novelty here is to extend
their analysis to the more general 2-dimensional dilaton gravity
theories while, at the same time, making it more systematic.  Our
method relies on the underlying symmetries of the particle motion
in a background geometry.  
In fact, the isometry of the background geometry 
we utilize in this paper to solve the motion exists for any gravity theory
for which the Birkhoff theorem holds, as the consideration
presented in section 2 shows.  Thus, our method
is applicable to a more general class of gravity theories than
those considered here.  For example, we can replace the dilaton
potential in (\ref{a1}) with an arbitrary function of the dilaton 
field $\phi$ and,
likewise, we can add $U(1)$ gauge fields to the action.   The 
Birkhoff theorem still holds under these types of generalizations
of (\ref{a1}) \cite{birkhoff} \cite{ab}.  
 
The details of our method are explained in section 2, along with 
a review of the background geometry of the gravity sector in 
consideration.  A new derivation of the background
geometry is presented in Appendix.  Although the most of the results 
in Appendix are already known \cite{kp}, they are 
tailored for our current purpose.
In section 3, we present the explicit expressions
of the particle trajectory for the CGHS model, the JT model, and
the 4-dimensional $s$-wave Einstein theory. We conclude this
paper by discussing various aspects of our results in section  4. 

\section{Background Geometry and Particle Trajectory}

In this section, we present the calculations leading to the analytic
expression of the particle 
motion in a 2-dimensional gravity theory, for which the
Birkhoff theorem holds.
For this purpose, we start by reviewing the background geometry of the
gravity sector in conformal gauge.  This also
serves to fix the notation.  The 
new derivation of the background geometry in conformal gauge, as
sketched in Appendix, motivates our calculations in section 2.2 and 
is instructive in its own right. 

\subsection{Background Geometry in Conformal Gauge}

The equations of motion for the background geometry are given
by varying the action $I_g$ in (\ref{a1}) with 
respect to the metric tensor
$g^{\alpha\beta}$ and the dilaton field $\phi$:
\begin{equation}
D_{\mu}D_{\nu} \Omega - g_{\mu \nu} D \cdot D \Omega
+ \frac{\gamma}{8} \left[g_{\mu \nu} \frac{(D \Omega )^2 }
{\Omega} - 2 \frac{D_{\mu } \Omega D_{\nu} \Omega}
{\Omega} \right] + \frac{\mu}{2}g_{\mu \nu} \Omega ^{1-\lambda}
=0,
\label{a5}
\end{equation}
\begin{equation}
R + \frac{\gamma}{4} \left[ \frac{(D \Omega )^2}{\Omega^2} - 2
\frac{D \cdot D \Omega}{\Omega} \right] + (1-\lambda)\mu\Omega^{-\lambda}
=0,
\label{a6}
\end{equation}
where $\Omega\equiv e^{-2\phi}$ and $D$ denotes the covariant
derivative.  In this paper, we choose to work in the conformal gauge where 
the metric takes the form
\begin{equation}
d s^2 =-e^{2\rho} dx^+ dx^- ,
\label{a8}
\end{equation}
where $x^{\pm}\equiv x^1 \pm x^0 $,
and the flat space metric is given by $g_{00}=- g_{11}=1$,
or $g_{+-}= g_{-+}=-1/2$\footnote{Under our signature choice,
$x^0$ is a time-like coordinate and $x^1$ is a space-like
coordinate.}.
Under this gauge choice, the equations of motion become
\begin{equation}
\partial_+ \partial_- ( \rho + \frac{\gamma}{8} 
\ln  \Omega ) + \frac{\gamma}{8}
\frac{\partial_+ \partial_- \Omega}{\Omega} + \frac{\mu(1-\lambda)}{8}
\Omega^{-\lambda }e^{2 \rho }=0,
\label{a10}
\end{equation}
\begin{equation}
\partial_+ \partial_- \Omega + \frac{\mu}{4}\Omega^{1-\lambda }
e^{2 \rho}=0 .
\label{a11}
\end{equation}
We should also impose 
the accompanying gauge constraints
\[
\frac{\delta I_g}{\delta g^{\pm \pm}} = 0 ,
\]
which we can read off from (\ref{a5}) as
\begin{equation}
\partial_{\pm}^2 \Omega 
- 2 \partial_{\pm} ( \rho + \frac{\gamma}{8} \ln \Omega )
\partial_{\pm} \Omega=0.
\label{a13}
\end{equation}
A specific set of solutions $(\rho , \Omega )$, obtained by solving
Eqs.(\ref{a10})-(\ref{a13}), corresponds to a specific background
geometry.  A method of getting the general solutions of the coupled 
partial differential equations in a local coordinate 
neighborhood is given in Appendix.  

     The general solutions in a local coordinate neighborhood depend on
two arbitrary chiral functions $f_{\pm} ( x^{\pm} )$.  
In terms of these two chiral functions, we define 
$X^{\pm} ( x^{\pm} )$
via $dx^{\pm} / dX^{\pm} = \exp ( f_{\pm} )$.  We further introduce
a variable 
\[ x = X^+ + \varepsilon_1 X^- \]
where $\varepsilon_1 = \pm 1 $.  We note that
if $\varepsilon_1 = +1 $, $x$ becomes a space-like 
variable, while $\varepsilon_1 = -1$
makes $x$ a time-like variable.  The dilaton field $\Omega 
= e^{ - 2 \phi }$ depends only on $x$ and is given implicitly by
\begin{equation}
\int \frac{d\Omega}{\varepsilon k\Omega^{a}-M/a}=\int dx,
\label{a25}
\end{equation}
where $M$ is a real parameter and we define
$a \equiv 2 - \lambda - \gamma / 4$.  Here the constant $k$ is
\begin{equation}
k=\Bigg\vert\frac{\mu e^{2\rho_0}}{4a}\Bigg\vert^{1/2}.
\label{a24-2}
\end{equation}
where $\rho_0$ is another real parameter.  The constant 
$\varepsilon = \pm 1$, depending on other parameters.  See Appendix
for more explanations on this point.  We can write down
the conformal factor $\rho$ as 
\begin{equation}
e^{2 \rho}= - {\rm sign}\left( \frac{\mu \varepsilon_1}{a} \right)
\Omega^{-\gamma /4} e^{ - ( f_+ + f_ - ) }
\left[ \Omega^{a} - \frac{M}{\varepsilon ka} \right]e^{2\rho_0}.
\label{a25-2}
\end{equation}
We note that in typical situations, such as the JT model, the
CGHS model, and the
$d$-dimensional $s$-wave Einstein theory, $\mu < 0$.  The presence
of two arbitrary chiral fields in the general solutions
simply represents the residual
gauge symmetry under our gauge choice, namely, the classical
conformal symmetry.  

     In Ref. \cite{birkhoff} and 
\cite{kp}, the same solutions as the above
were obtained for $\varepsilon_1 = + 1$ case, i.e., when the
dilaton field is space-like, under a different gauge choice.
The solutions for the CGHS case represent the dilatonic black
hole and the linear dilaton vacuum \cite{wadia}.  In case of the JT theory,
the local solutions were used to construct a cylindrical
geometry \cite{jt}.  
For $\varepsilon_1 = -1$ case, we have a situation where the 
dilaton depends only on
a time-like variable.  Our further analysis is valid for
every solution shown in the above, regardless of whether we have
a static background geometry or not.

\subsection{Particle Motion in a Background Geometry}

We now consider the motion of a point particle in a generic
background geometry described by (\ref{a25}) and (\ref{a25-2}).
The particle trajectory on a generic background geometry 
is determined by the geodesic equation from (\ref{a2})
\begin{equation}
\frac{d^2 x^{\mu}}{d \tau^2}=-\Gamma^{\mu}_{\nu \lambda}
{dx^{\nu}\over d\tau }
{dx^{\lambda}\over d\tau },
\label{a7}
\end{equation}
where the usual Christoffel symbols are introduced as 
$\Gamma^{\mu}_{\nu \lambda} = 
{1\over 2}g^{\mu\rho}({\partial}_{\nu}g_{\lambda\rho}
+{\partial}_{\lambda}g_{\nu\rho}-{\partial}_{\rho}g_{\nu\lambda})$.
As in section 2.1, we choose to describe the particle motion in the
conformal gauge.  Then, the equations of motion (\ref{a7}) become
\begin{equation}
\frac{d^2x^{\pm}}{d\tau^2}=-2\partial_{\pm} \rho \left(\frac{dx^{\pm}}
{d\tau}\right)^2.
\label{a26}
\end{equation}
We have a further freedom to choose a particular set 
of conformal coordinates,
thereby fixing the residual conformal symmetry.  It turns out that 
the choice of $(X^+ (x^+ ) , X^- (x^- ) )$ fields as our 
conformal coordinates provides a simplest description of the particle
motion.  Thus, we rewrite (\ref{a26}) as
\begin{equation}
\frac{d^2X^{\pm}}{d\tau^2}=-2
{\partial \tilde{\rho}\over \partial X^{\pm}}
\left(\frac{dX^{\pm}}{d\tau}\right)^2,
\label{a27}
\end{equation}
where the new conformal factor can be calculated to be
$\tilde{\rho}(x) = \rho +(f_{+}+f_{-})/2$.  The crucial property
of this gauge fixing is that the function $\tilde{\rho}$ now depends
on $(X^+ , X^- )$ only through the combination 
$x = X^+ + \varepsilon_1 X^-$ (see
Eq.(\ref{a25-2}) and Appendix).  This is a consequence of the
Birkhoff theorem, which ensures that the general solutions of our
gravity sector look locally isomorphic to
static solutions, under a particular coordinate choice.
For a gravity theory where the Birkhoff theorem holds, we
can straightforwardly determine the particle motion using
this property\footnote{Thus,
the analysis here applies to more general 
actions than the class given in (\ref{a1}), as long as the Birkhoff
theorem holds.  In the context of
the CGHS model, for example, we can consider a theory with
the loop corrections from string theory.}.

  To integrate the system described by (\ref{a27}) reducing it to
a first-order system, we need two symmetries.  One of these
symmetries is obvious; our particle action is invariant under
the affine transformation of the proper time parameter $\tau$.
The translation transformation of the proper time parameter,
$ \delta \tau  = \epsilon_1$, where $\epsilon_1$ is a constant, is thus
a symmetry.  We can construct its Noether charge
\begin{equation}
  c_0 =e^{2\tilde{\rho}} \dot{X}^+ \dot{X}^-  ,
\label{a28-1}
\end{equation}
where the overdot represents a differentiation with respect to $\tau$.
Since we are interested in the motion of a massive particle,
we set $c_0 = -1$.  Eq.(\ref{a28-1}) then becomes 
the mass-shell condition for a massive relativistic
particle.  Another symmetry of the particle motion is due to the
property of the background geometry.
The single variable dependence of the conformal factor 
$\tilde{\rho} (x)$ permits us to easily find one isometry of the 
background geometry.  Namely, under the transformation
\[
\delta X^+ ={\epsilon}_2 \ ,\delta X^- = - \varepsilon_1{\epsilon}_2 , 
\]
where $\epsilon_2$ is a constant, the background metric is
invariant.  In case of a flat, static background geometry written in
terms of tortoise coordinates, it reduces to the
Lorentz transformations.  This isometry of the background geometry
induces another Noether charge
\begin{equation}
c_1 = e^{2\tilde{\rho}} (\dot{X}^- - \varepsilon_1 \dot{X}^+) 
\label{a28-2}
\end{equation}
for the motion of the particle. We have, thus, reduced our problem
to a first-order system, Eqs.(\ref{a28-1}) and (\ref{a28-2}).

   Further integration of the first-order system is also
straightforward once we consider the properties of the background
geometry.  By solving (\ref{a28-1}) and (\ref{a28-2}), we 
have 
\begin{equation}
\dot{X}^+ = \frac{-c_1 \pm \sqrt{c_1^2 -4\varepsilon_1 e^{2\tilde{\rho}}}}
{2\varepsilon_1 e^{2\tilde{\rho}}}~,~~~\dot{X}^- = 
\frac{c_1 \pm \sqrt{c_1^2 -4\varepsilon_1
e^{2\tilde{\rho}}}}{2 e^{2\tilde{\rho}}}. 
\label{temp}
\end{equation}
Noting that 
$x = X^+ + \varepsilon_1 X^-$ and $y = X^+ - \varepsilon_1 X^-$ are 
mutually orthogonal coordinates of the background geometry, we
can rewrite (\ref{temp}) as
\begin{equation}
\dot{x}=\pm \varepsilon_1 
e^{-2\tilde{\rho}(x)} \sqrt{c_1^2-4\varepsilon_1 e^{2\tilde{\rho}(x)}}
\label{a29}
\end{equation}
and
\begin{equation}
\dot{y}=-c_1 \varepsilon_1 e^{-2\tilde{\rho}(x)} .
\label{a30}
\end{equation}
We now have explicitly decoupled differential equations.  
To get the parametrized form of the particle trajectory, we first
solve for $x(\tau)$ using (\ref{a29}).  Then, $y(\tau)$ can be
obtained from (\ref{a30}). The particle trajectory can also be
written as
\begin{equation}
\int dy = \pm \int \frac{c_1dx}{\sqrt{c_1^2-
4\varepsilon_1e^{2\tilde{\rho}(x)}}}
\label{traj}
\end{equation}
from (\ref{a29}) and (\ref{a30}).  

\section{Explicit Examples}

We evaluate the integrations in (\ref{traj}) for
the JT model, the CGHS model and the 
4-dimensional $s$-wave Einstein theory, 
to get explicit expressions for
the particle trajectory.  Our JT theory results
are compared to those of 
Ref. \cite{hksy}\footnote{We actually find, in that case,
the solutions presented in \cite{hksy} are not complete,
since we find another class of trajectories.}.  We will 
discuss the CGHS model and the $d$-dimensional $s$-wave 
Einstein gravity together.

\subsection{Jackiw-Teitelboim Model}

This model corresponds to the choice of $\gamma =\lambda =0$ 
and $\mu=-\Lambda/2$ in (\ref{a1}).
The $\Omega$ field can be calculated from Eq. (\ref{a25}) to be
\begin{equation}
\Omega=-\varepsilon \sqrt{\frac{M}{2\varepsilon k}} 
\tanh{ \left[ \sqrt{\frac{kM}{2\varepsilon }}(x-x_0) \right]}
\label{a35-1}
\end{equation}
where $k=|\Lambda e^{2\rho_0}/16|^{1/2}$ and $x_0$ is the constant of
integration.  The conformal factor $\tilde{\rho}$ is 
given from Eq. (\ref{a25-2}) as
\begin{equation}
e^{2\tilde{\rho}}=\frac{|M|e^{2\rho_0}/(2k)}{\cosh^2
\sqrt{\frac{kM}{2\varepsilon }}(x-x_0)}.
\label{a35-2}
\end{equation}
We note that hyperbolic functions in the above Eqs. (\ref{a35-1})
and (\ref{a35-2}) become trigonometric functions 
in case of $M/\varepsilon <0$
using $\tanh ix=i\tan x$ and $\cosh ix=\cos x$.
Using $e^{2\rho}=e^{2\tilde{\rho}-f_+-f_-}$, the metric becomes
\begin{equation}
e^{2\rho}={\rm{sign}}\left(
\frac{\varepsilon  \Lambda}{M \varepsilon_1} \right)
\frac{2\partial_+A\partial_-B}{\left(1+\Lambda
AB/8 \right)^2}
\label{a35-3}
\end{equation}
where $A(x^+)=(8/\Lambda)^{1/2}e^{\sqrt{2kM/\varepsilon }(X^+-x_0/2)}$ and
$B(x^-)=(8/\Lambda)^{1/2}
e^{\sqrt{2kM/\varepsilon }(\varepsilon_1X^--x_0/2)}$.
The pre-factor $\rm{sign} (\varepsilon \Lambda / M \varepsilon_1 ) $ of  
the right hand side of Eq.(\ref{a35-3}) can be determined 
from the table (\ref{table1})
by multiplying the first, the second and the fourth column.  
We find there are two cases
when this pre-factor is positive, i.e., the third and the last row of
the table (\ref{table1}).  In these cases, however, the range 
of $\Omega'$, which is given by $\Omega' =-(M/2) {\rm sech}^2
\sqrt{(kM/2\varepsilon )}(x-x_0)$, is inconsistent with the range 
given in the table (\ref{table1}).  For consistency,  we have to  consider
only the case of
${\rm sign} (\varepsilon  \Lambda/M \varepsilon_1)=-1$.  This result is
consistent with that of Ref. \cite{hksy}.

When $\Lambda > 0$, $M > 0$ and $\varepsilon_1=-1$, we get
$\varepsilon =+1$ from the table (\ref{table1}).
Then, the field $\Omega$ is calculated from Eq. (\ref{a35-1}) to be
\begin{equation}
\Omega=-\sqrt{\frac{M}{2k}} \tanh{ \left[ \sqrt{\frac{kM}{2}}(x-x_0)
\right]}.
\label{a36}
\end{equation}
We can easily verify that $\Omega$ satisfies the condition,
$-M/2<\Omega'<0$, in the table (\ref{table1}). We get $\tilde{\rho}$ from 
Eq. (\ref{a35-2})
\begin{equation}
e^{2\tilde{\rho}}=\frac{Me^{2\rho_0}/(2k)}{\cosh^2
\sqrt{\frac{kM}{2}}(x-x_0)}
\label{a37}
\end{equation}
and the particle trajectory is given by Eq. (\ref{traj})
\begin{equation}
\sinh{\sqrt{\frac{kM}{2}}(x-x_0)}=\pm \sqrt{1+\frac{2Me^{\rho_0}}{kc_1^2}}
\sinh{\sqrt{\frac{kM}{2}}(y-y_0)}  .
\label{a39}
\end{equation}
This is the result obtained in Ref. \cite{hksy}.   Here, we find
another class of 
trajectory not reported there.
When $\Lambda >0$, $M > 0$ and $\varepsilon_1=+1$, from 
the table (\ref{table1}),
we have $\varepsilon =-1$ and $\Omega'<-M/2$.   The results of
our calculations
in this case are  
\begin{equation}
\Omega=-\sqrt{\frac{M}{2k}} \tan{ \left[ \sqrt{\frac{kM}{2}}(x-x_0)
\right]},
\label{a40}
\end{equation}
\begin{equation}
e^{2\tilde{\rho}}=\frac{Me^{2\rho_0}/(2k)}{\cos^2
\sqrt{\frac{kM}{2}}(x-x_0)},
\label{a41}
\end{equation}
and
\begin{equation}
\sin{\sqrt{\frac{kM}{2}}(x-x_0)}=\pm \sqrt{1-\frac{2Me^{\rho_0}}{kc_1^2}}
\sin{\sqrt{\frac{kM}{2}}(y-y_0)} . 
\label{a42}
\end{equation}

\subsection{CGHS and 4-Dimensional Model}

We investigate the particle motion in the CGHS model and
the spherically symmetric 4-dimensional Einstein gravity.
At the classical level, we can treat them
together since the CGHS model ($a=1$) is the same as the $d= \infty$ limit 
of the spherically symmetric $d$-dimensional gravity where 
$a=(d-3)/(d-2)$.
To simplify the presentation, we regard $\tilde{\rho}$ as a function 
of $x$ through
$\Omega$.  We have
\begin{equation}
e^{2\tilde{\rho}}=-{\rm sign}\left(\frac{\mu \varepsilon_1}{a} \right)
e^{2\rho_0}\left[1-\frac{M}{\varepsilon ka} \Omega^{-a}
 \right]
\label{a45}
\end{equation}
from Eq.(\ref{a25-2}).  The particle trajectory 
given by Eq.(\ref{traj}) becomes
\begin{equation}
\int dy = \pm\frac{1}{M}
\int \frac{dt}{t\sqrt{1-Ct}}\left[\frac{\varepsilon ka}{M}\left({\rm sign}
\left(\frac{\mu}{a} \right) t +1 \right) \right]^{-1/a}
\label{a46},
\end{equation}
where $t=-{\rm sign}(\mu/a)[1-M(\varepsilon ka\Omega^{a})^{-1}]$
and $C=4e^{2 \rho_0}/c_1^2$. 
Now we consider the particle trajectory when
$M>0$, $\varepsilon_1=+1$ and $\mu<0$, since this choice represents
the black hole solutions, as can be seen in Ref. \cite{kp}.  From the 
table (\ref{table1}), then, we have $\varepsilon =+1$ and $0<\Omega'$.
Note that the variable $t(\Omega)$ satisfies $0<t<1$, which follows from
$0 < \Omega'$.  We note $t\sim 0$ and $t \sim 1$ represent  the 
space-time region near the black hole horizon and the region close to
the asymptotic infinity, respectively.

In case of the CGHS model,  we have $a=1$.   From
Eq.(\ref{a25}), we get
\begin{equation}
\Omega=\frac{M}{ k}  + e^{\varepsilon k(x-x_0)} ,
\label{a42-1}
\end{equation}
the dilatonic black hole.
The Eq. (\ref{a25-2}) gives us $\tilde{\rho}$ as
\begin{equation}
\tilde{\rho}=\frac{1}{2} k(x-x_0) - \frac{1}{2} \ln \Omega + \rho_0 .
\label{a42-2}
\end{equation}
By integrating Eq.(\ref{a46}), we get
the particle trajectory as
\begin{equation}
y-y_0=\pm \frac{1}{k} \left[ -2 \tanh^{-1} \sqrt{1-Ct} + \frac{2}
{\sqrt{1-C}} \tanh^{-1} \sqrt{\frac{1-Ct}{1-C}} \right] .
\label{a48}
\end{equation}

In case of the 4-dimensional Einstein theory, we have the 
parameter of $a= 1/2$.
We get from Eq.(\ref{a25})
\begin{equation}
2\Omega^{1/2}+\frac{4M}{k} \ln \Bigg\vert \Omega^{1/2}-\frac{2M}{k}
\Bigg\vert =  k(x-x_0) , 
\label{a44}
\end{equation}
where  $k=|\mu e^{2\rho_0}/2|^{1/2}$ and $x_0$ is a constant of
integration.  The conformal factor $\tilde{\rho}$ can be calculated 
from Eq.(\ref{a44}) and
Eq. (\ref{a25-2}).
Using Eq. (\ref{a46}), the particle trajectory
can be calculated to be
\[ y - y_0 = \]
\begin{equation}
\pm \frac{4M}{k^2} \left[ -2 \tanh^{-1} \sqrt{1-Ct} 
+ \frac{2-3C}{(1-C)^{3/2}} \tanh^{-1} \sqrt{\frac{1-Ct}{1-C}}
+\frac{\sqrt{1-Ct}}{(1-C)(1-t)} \right] .
\end{equation}

\section{Discussions}
We presented a method to determine the motion of a test particle
in a background geometry of 2-dimensional gravity theories
satisfying the Birkhoff theorem.   The existence of the 
isometry of the background 
geometry that  played a crucial role in our method is not specific to a 
certain choice of the gravity sector, but 
it is the general property shared by 
the general 2-dimensional dilaton gravity theory.   
It is interesting to note
that this type of unified description is possible 
for a large class of theories.
We expect the local  solutions we get in this paper can be useful building
blocks to form non-trivial global structures of the space-time and
the resulting particle motion.  The study of possible global constructions 
can further illuminate the similarities and differences among the 
theories we consider in this paper.  

Given our results, there are some further issues that can be addressed.
The detailed structure of the isometries of the background geometry is
interesting in itself.  In flat space-time geometry, our gravity sector's
isometries form the 2-dimensional Poincar\'e algebra.  
As the curvature effects
creep in, this algebra gets deformed.  The isometry we utilized in this
work is one part of that algebra.   
From our solutions, we can possibly uncover 
the detailed structure of the deformed algebra. 

The quantization of a massive particle in a background 
geometry, including black hole geometries and other 
non-trivial global geometries,
is an issue of great importance.  The general solutions for the 
particle trajectories
we obtained in this paper can be a useful starting 
point for such investigation.
By calculating the symplectic structure on the space 
of all classical solutions,
we can see the structure of the quantum phase space and proceed to the 
quantization of our classical problem.  
Our work in this direction is in progress,
for the purpose of providing a simple quantum mechanical system that can
capture essential features of the quantum black hole physics.  


\appendix
\begin{flushleft}{\Large \bf Appendix : Derivation of Background
Geometry}
\end{flushleft}

We obtain the general (local) solution $\rho$
and $\Omega$ of (\ref{a10}) and (\ref{a11}) under the 
constraints (\ref{a13}).  For convenience, we introduce
$\bar{\rho} = \rho + (\gamma \ln \Omega ) / 8$.  
By integrating the gauge constraints (\ref{a13}), we get
\begin{equation}
\ln{|\partial_{\pm} \Omega |} = 2 \bar{\rho} + f_{\mp}(x^{\mp}) ,
\label{a14}
\end{equation}
where $f_+(x^+)$ and $f_-(x^-)$ are arbitrary functions of $x^+$ and $x^-$,
respectively. By taking the difference of two equations in (\ref{a14}),
we get
\begin{equation}
\left| \frac{\partial \Omega}{\partial X^+}  \right|
=  \left| \frac{\partial \Omega}{\partial X^-} \right|,
\label{a15}
\end{equation}
where $X^{\pm}$ is defined by $dx^{\pm}/dX^{\pm}\equiv e^{f_{\pm}}$.
This means that $\Omega$ is a function of only one variable
$x\equiv X^+ + \varepsilon_1 X^-$ where $\varepsilon_1=\pm 1$, 
since (\ref{a15}) implies
that $\Omega=\Omega(X^+ \pm X^-)$.
Furthermore, since $\partial_+ \partial_- \Omega
=\varepsilon_1 e^{-f_+-f_-}\Omega''$
(where the prime denotes the differentiation with respect to $x$),
we find, from (\ref{a11}), that $\hat{\rho}(x)\equiv
\bar{\rho}+(f_++f_-)/2$ is also a function of only $x$.
Using this property we can rewrite the equations of motion
(\ref{a10}) and (\ref{a11}) as
\begin{equation}
{\hat{\rho}}''+ \frac{\mu}{8}\varepsilon_1(1-\lambda-\gamma/4)
\Omega^{-\lambda-\gamma/4}e^{2\hat{\rho}} =0,
\label{a17}
\end{equation}
\begin{equation}
{\Omega}''+\frac{\mu}{4}\varepsilon_1 \Omega^{1-\lambda-\gamma/4}
e^{2\hat{\rho}}=0,
\label{a18}
\end{equation}
and the gauge constraint (\ref{a13}) as 
\begin{equation}
{\Omega}''-2{\hat{\rho}}' \, {\Omega}'=0.
\label{a19}
\end{equation}
Thus, our system of partial differential equations effectively
becomes a system of ordinary differential equations.  
This provides an alternative proof
of the Birkhoff theorem, originally proved for this type
of theories in \cite{birkhoff} and \cite{ab}.

General solutions of (\ref{a17}) and
(\ref{a18}) under the gauge constraint (\ref{a19}) can be obtained 
by following the method used in Ref. \cite{kp}.
First, we find an
effective action that produces (\ref{a17}) and (\ref{a18}) 
\begin{equation}
I_{eff}=\int dx \left( \hat{\rho}' \Omega'-\frac{\mu}{8}\varepsilon_1
e^{2\hat{\rho}} \Omega^{1-\lambda-\gamma/4} \right).
\label{a20-1}
\end{equation}
Then we find the symmetries of the effective action
to construct the Noether charges corresponding to them.
We note that the effective action (\ref{a20-1}) is invariant under
\begin{eqnarray}
& &\delta x=\epsilon_1,
\label{a20-2}\\
& &
\delta x =\epsilon_2 x,\
\delta \Omega=\epsilon_2 \Omega,\
\delta \hat{\rho}=-\frac{1}{2} \epsilon_2
\left(2-\lambda-\frac{\gamma}{4} \right),
\label{a20-3}
\end{eqnarray}
producing two Noether charges
\begin{equation}
q = {\hat{\rho}}' {\Omega}' +
\frac{\mu}{8}\varepsilon_1e^{2\hat{\rho}} \Omega^{1-\lambda-\gamma/4},
\label{a20}
\end{equation}
\begin{equation}
\frac{M}{2}=- q x -\frac{1}{2} {\Omega}' \left(2-\lambda-
\frac{\gamma}{4}\right)+ {\hat{\rho}}' \Omega.
\label{a21}
\end{equation}
We can alternatively integrate the original second-order system, 
Eqs.(\ref{a17}) and (\ref{a18}), by expressing them
as the conservation laws of the Noether charges, to obtain the
first-order system shown above.
Now, the gauge constraint (\ref{a19}) fixes $q=0$.
With $q = 0$, by integrating (\ref{a21}), we get
\begin{equation}
\hat{\rho}=\frac{M}{2} \int \frac{dx}{\Omega} + \frac{1}{2}a \ln{\Omega}
 + \rho_0,
\label{a22}
\end{equation}
where $a \equiv 2-\lambda-\gamma/4$
and $\rho_0$ is a constant
of integration. Plugging (\ref{a22}) into (\ref{a20}), we get a
decoupled equation for $\Omega$ that does not contain $\hat{\rho}$:
\begin{equation}
M \Omega'+a\left(\Omega' \right)^2=-\frac{\mu}{4}
\varepsilon_1 \Omega^{2a} 
\exp{\left(M \int \frac{dx}{\Omega} +2\rho_0 \right)}.
\label{a22-1}
\end{equation}
We can solve this equation as follows.  We differentiate it with respect 
to $x$ to get
\begin{equation}
\left[M+2a\Omega' \right] \left[ \Omega''
- \left( M+a\Omega' \right)
\Omega'/\Omega \right]=0.
\label{a23}
\end{equation}
When the first factor of the left hand side of Eq. (\ref{a23}) vanishes,
we have $\Omega''=0$, and then $\Omega=0$ from (\ref{a18}).
That means $\phi =\infty$ everywhere, and we discard this case.
When the second factor vanishes, we have 
\begin{equation}
\frac{d^2 \Omega}{dx^2}
= \left[ M+a\frac{d \Omega}{dx} \right]
\frac{d \Omega}{dx} \frac{1}{\Omega} ,
\label{a23-4}
\end{equation}
which is integrated to yield
\begin{equation}
k \Omega^{a}= \Bigg\vert \frac{M}{a} + \frac{d \Omega}{dx}
\Bigg\vert    = \varepsilon  \left(  \frac{M}{a} 
+ \frac{d\Omega}{dx} \right) ,
\label{a24}
\end{equation}
where $k$ is a positive constant of integration 
and the constant $\varepsilon = +1$
or $-1$ that comes in when we remove the absolute sign.
We must fix the constant $k$ since both (\ref{a24}) and
the original equation (\ref{a22-1}) are
first order differential equations.
By integrating (\ref{a23-4}), we get
\begin{equation}
\int \frac{M}{\Omega}dx=\ln{\Bigg\vert\frac{d\Omega/dx}{M/a+
d\Omega/dx} \Bigg\vert}.
\label{a24-1}
\end{equation}
By plugging (\ref{a24}) and (\ref{a24-1})
into (\ref{a22-1}), we find that $k$ is fixed as
\begin{equation}
k=\Bigg\vert\frac{\mu e^{2\rho_0}}{4a}\Bigg\vert^{1/2}.
\end{equation}
By integrating (\ref{a24}), we obtain the solution for $\Omega$:
\begin{equation}
\int \frac{d\Omega}{\varepsilon k\Omega^{a}-M/a}=\int dx.
\label{redun}
\end{equation}
  From (\ref{a22})
we obtain the solution for $\hat{\rho}$:
\begin{equation}
\hat{\rho}=\frac{1}{2}\ln{\Bigg\vert \varepsilon \Omega^{a}-\frac{M}{ka}
\Bigg\vert}+\rho_0.
\end{equation} 
Using Eqs. (\ref{a22-1}) and (\ref{a24-1}), the sign of
$\Omega'=\varepsilon k\Omega^{a}-M/a$ can be calculated as ${\rm sign}
(\Omega')=-{\rm sign}(\mu \varepsilon_1 \varepsilon /a)$, from 
which we get
\begin{equation}
e^{2\hat{\rho}}={\rm sign}\left( \frac{\mu \varepsilon_1}{a} \right)
\left[ \frac{M}{\varepsilon k a} -\Omega^{a} \right]e^{2\rho_0}.
\label{a25-1}
\end{equation}
This equation, along with (\ref{redun}), provides us with the
desired results, namely, Eqs. (\ref{a25}) and (\ref{a25-2}). 
We determine the value of $\varepsilon$ as follows; from the requirement
that the exponential function in Eq. (\ref{a22-1}) should be positive, we 
determine the range of $\Omega^{\prime}$.
Then, using the range, we choose the value of $\varepsilon$ that makes
the right hand side of (\ref{a24}) positive.  To summarize, we have
\begin{equation}
\begin{tabular}{|c|c|c|c|} \hline
$\mu \varepsilon_1  /a $ & $M/a$ &
range of $\Omega'$ & $\varepsilon$ \\ \hline
+   & +   & $-M/a< \Omega' <0$ & +1   \\
+   & $-$ & $ 0<\Omega'<-M/a$  & $-1$ \\
$-$ & +   & $0<\Omega'$            & +1   \\
$-$ & +   & $\Omega'<-M/a$     & $-1$ \\
$-$ & $-$ & $-M/a<\Omega'$     & +1   \\
$-$ & $-$ & $\Omega' <0$           &$-1$  \\ \hline
\end{tabular}
\label{table1}
\end{equation}
where + and $-$ indicate the signs.  For example, to have a 
description of the 4-dimensional Schwarzschild geometry, where
$a = 1/2$, $\mu = -2$ and the black hole mass 
$M > 0$, we need the first and the third row results.
For the inside of the black hole, the dilaton field depends
on the time-like variable (therefore $\varepsilon_1 = -1$) and the
first row is the case.  For the outside of the black hole,
the dilaton field depends on the space-like variable
(therefore $\varepsilon_1 = +1 $) and we have the third row.

\end{document}